\documentclass[lettersize,journal]{IEEEtran}

\usepackage{amsmath,amsfonts}
\usepackage{array}
\usepackage[tight,footnotesize]{subfigure}
\usepackage{textcomp}
\usepackage{stfloats}
\usepackage{url}
\makeatletter

\usepackage{hyperref}
\makeatother
\usepackage{verbatim}
\usepackage{graphicx}
\usepackage{cite}

\usepackage{graphicx}
\graphicspath{{image/}}
\DeclareGraphicsExtensions{.pdf}
\usepackage{multirow}
\usepackage{booktabs}
\usepackage{colortbl}
\usepackage{xcolor}
\usepackage{xspace}
\usepackage{enumitem}
\usepackage{framed}
\usepackage[marginal]{footmisc}
\usepackage[ruled,linesnumbered]{algorithm2e}

\newcommand{\code}[1]{{\fontfamily{cmtt}\fontseries{m}\fontshape{n}\selectfont\small{#1}}}

\newcommand{\tab}{\hspace*{1em}}
\newcommand{\system}{Esuer\xspace}

\definecolor{mygray}{gray}{.9}

\begin{document}

\title{Building Reuse-Sensitive Control Flow Graphs (CFGs) for EVM Bytecode}

\author{Dingding Wang, Jianting He, Yizheng Yang, Lei Wu, Rui Chang, Yajin Zhou\thanks{
Dingding Wang, Yizheng Yang, Lei Wu, Rui Chang, Yajin Zhou are with the School of Cyber Science and Technology, Zhejiang University, Hangzhou, China. Jianting He is with the Blocksec Team, Hangzhou, China.
\\Email: dingdingwang@zju.edu.cn, jthe@blocksec.com, \{yizhengyang, lei\_wu, crix1021, yajin\_zhou\}@zju.edu.cn.
\\Corresponding author: Yajin Zhou.}}

\markboth{Journal of \LaTeX\ Class Files,~Vol.~14, No.~8, August~2021}%
{Shell \MakeLowercase{\textit{et al.}}: A Sample Article Using IEEEtran.cls for IEEE Journals}


\maketitle

\begin{abstract}

The emergence of smart contracts brings security risks, exposing users to the threat of losing valuable cryptocurrencies, underscoring the urgency of meticulous scrutiny.
Nevertheless, the static analysis of smart contracts in EVM bytecode faces obstacles due to flawed primitives resulting from code reuse introduced by compilers. Code reuse, a phenomenon where identical code executes in diverse contexts, engenders semantic ambiguities and redundant control-flow dependencies within reuse-insensitive CFGs.

This work delves into the exploration of code reuse within EVM bytecode, outlining prevalent reuse patterns, and introducing \system, a tool that dynamically identifies code reuse when constructing CFGs.
Leveraging taint analysis to dynamically identify reuse contexts, \system identifies code reuse by comparing multiple contexts for a basic block and replicates reused code for a reuse-sensitive CFG.
Evaluation involving 10,000 prevalent smart contracts, compared with six leading tools, demonstrates \system’s ability to notably refine CFG precision. It achieves an execution trace coverage of 99.94\% and an F1-score of 97.02\% for accurate identification of reused code. Furthermore, \system attains a success rate of 99.25\%, with an average execution time of 1.06 seconds, outpacing tools generating reuse-insensitive CFGs. \system's efficacy in assisting identifying vulnerabilities such as tx.origin and reentrancy vulnerabilities, achieving F1-scores of 99.97\% and 99.67\%, respectively.

\end{abstract}

\begin{IEEEkeywords}
Control Flow Graph, EVM , bytecode
\end{IEEEkeywords}

\section{Introduction}
\label{sec:intro}

Nowadays, smart contracts have gained widespread popularity, involving billions of USD every day.
At the same time, smart contracts are error-prone~\cite{vul1,vul2,vul3} or malicious~\cite{attack1,attack2}, resulting in a potentially high financial risk to their users. For instance, in 2022, attacks on vulnerable smart contracts caused a loss of about 4 billion USD~\cite{loss}.

To mitigate the threat, static analysis tools have been developed to detect \textit{benign but vulnerable smart contracts} and \textit{malicious contracts}.
Several tools~\cite{brent2018vandal,rattle,contro2021ethersolve,pasqua2023enhancing,octopus,luu_making_2016,grech2018madmax,albert2018ethir} operate on EVM bytecode, which represents the compiled form of smart contracts.
The absence of verified source code constitutes a key enabling factor for these tools.
Studies~\cite{zhou2018erays,liao2022large} indicate that approximately 77.3\% of deployed smart contracts lack verified source code on Etherscan~\cite{etherscan}.
Significantly, the source code for almost all malicious smart contracts remains inaccessible.
Furthermore, even among open-source smart contracts, about 11.9\% contain inline assembly code that poses challenges for source-level analysis tools.

\noindent\textbf{Limitations of existing tools}\tab
However, the efficacy of static analysis tools is significantly hindered by the limitations within CFGs due to code reuse in EVM bytecode.
Code reuse occurs when two distinct execution paths share basic blocks (BBs) in bytecode, despite having disparate contexts.
It is generated by compilers to compress smart contract size, which can save transaction fees for deploying.
However, this practice poses intricate challenges for static analysis tools.
Existing tools neglect code reuse and generate reuse-insensitive CFGs, which contain an explosion of control-flow dependency because of disparate execution contexts for the same BB.
While certain decompilers~\cite{grech2019gigahorse,grech2022elipmoc,zhou2018erays} work on recovering functions, which eases the explosion, they cannot eliminate code reuse in generated CFGs.
The limitations on the reuse-insensitive CFG for EVM bytecode are explained in detail in \S\ref{sec:motivating}.

\noindent\textbf{Our observations and solution}\tab
Code reuse presents complex implications for static analysis.
Despite mentions in previous work~\cite{grech2022elipmoc,zhou2018erays} regarding compiler optimizations in code reuse, a comprehensive investigation is absent.
Consequently, we systematically study the root cause and impacts of code reuse to enable reuse-sensitive analysis.
In summary, we extract eight prevalent code reuse patterns and elucidate two challenges in reuse-sensitive analysis in \S\ref{sec:BB_reuse}.



In response to the challenges, we introduce \system, a tool tailored to generate reuse-sensitive CFGs for EVM bytecode.
\system's workflow involves bytecode preprocessing for BB generation (\S\ref{subsec:pre-process}) and iterative construction of reuse-sensitive CFGs (\S\ref{subsec:construct_cfg}).
The construction begins with an empty stack at the first block, with each iteration handling a BB and finding its successors, progressing through successors until all control-flow transfers are addressed.
Each iteration comprises three steps, dynamically completing the reuse context for each BB.
With insights into code reuse, \system utilizes jump operands on the stack to determine the reuse context, where distinct contexts for the same BB signify code reuse.
Specifically, \system captures stack state snapshots (\S\ref{subsubsec:snapshot_generation}), taints reuse context based on snapshots (\S\ref{subsubsec:update_context}), and finds a non-reused successor or generates a clone if necessary (\S\ref{subsubsec:reuse_handler}).
Detection of reuse relies on comparing reuse contexts across iterations.

We conduct a comprehensive evaluation of \system by comparing it with six SoA tools.
The assessments in precision and performance utilize 10,000 prevalent smart contracts as dataset.
The results show that \system achieves the best precision, with a execution trace coverage of 99.94\%, the fewest infeasible paths, and a F1-score of 97.02\% in code reuse identification correctness.
In addition, \system achieves the highest success rate at 99.25\%, with an average execution time of 1.06 seconds, surpassing most tools that construct reuse-insensitive CFGs, while demonstrating robust performance across smart contract sizes.
Furthermore, we evaluate the effectiveness of \system in downstream static analysis with SolidiFi dataset~\cite{ghaleb2020effective}.
Detectors based on \system achieve a F1-score of 99.97\% to detect tx.origin vulnerabilities and a F1-score
of 99.67\% to detect reentrancy vulnerabilities.
The outperformance of \system in terms of precision, performance, and effectiveness provides strong evidence of its reliability and potential.

In summary, our work makes the following contributions.
\begin{itemize}[leftmargin=*]
    \setlength{\itemsep}{0pt}
    \setlength{\parsep}{0pt}
    \setlength{\parskip}{0pt}
    \item We are the first to study code reuse introduced by compilers, proposing insights on common code reuse patterns and challenges to static analysis of EVM bytecode.
    \item We propose a method to address the code reuse issue, and implement \system to construct reuse-sensitive CFGs.
    \item We evaluate \system with six SoA tools, demonstrating its superiority in precision, performance, and effectiveness.
\end{itemize}

\section{Background}
\label{Sec:Background}

\subsection{EVM Bytecode}
\label{subsec:EVM}

Ethereum Virtual Machine (EVM), is the defacto standard to execute smart contracts, used by many popular programmable blockchains, such as Ethereum~\cite{wood2014ethereum}, Binance Smart Chain (BSC)~\cite{bsc}, and Polygon~\cite{polygon}.
Its assembly language, EVM bytecode, is a stack-based low-level intermediate representation where operands are obtained from stack.
There are some special instructions in the EVM bytecode related to our work.
First, only the \code{PUSH} instructions~\footnote{We use \code{PUSH} instructions to denote the instructions with opcodes from \code{0x60} to \code{0x7F}. They are used to push different lengths of data on the stack.} have data encoded in instructions. 
All other instructions contain only one byte of opcode.
Second, the instruction \code{JUMPDEST} (whose opcode is \code{0x5b}) is used to denote the addresses of valid jump targets.
All jumps to other instructions except \code{JUMPDEST} will be reverted by EVM.
Third, the \code{JUMP} instructions (including unconditional \code{JUMP} and conditional \code{JUMPI}) uses the top value in the stack as jump targets, instead of encoding jump targets in instructions like x86 or JVM bytecode.

\subsection{CFG Construction}
\label{subsec:static_analysis}


CFG is essential for most static techniques, greatly influencing their effectiveness.
CFG is a directed graph, in which nodes are BBs and edges are control-flow transfers.
BB is a sequence of instructions with one entry and one exit.
There are two special types of nodes in CFG: join node and branch node.
Join node is the node with multiple predecessors, and branch node is the node with multiple successors.

CFG recovery has been widely discussed in prior work~\cite{kinder2008jakstab,kruegel2004static,schwarz2002disassembly,xu2009constructing}.
Two primary strategies for constructing CFG are discussed: linear sweep and recursive traversal~\cite{schwarz2002disassembly}.
Linear sweep is to linearly disassemble and split BBs, extract jump targets from jump instructions, and add the new edge to the CFG.
The whole CFG can be constructed in a single pass.
However, this strategy cannot handle interleaved code and data sections (e.g., in ARM assembly language).
Recursive traversal is to disassemble and analyze a BB, identify its successors, connect them, and repeat the analysis recursively for all successors until no new BBs are identified.

Existing approaches to construct CFGs for EVM bytecode take the recursive traversal strategy~\cite{brent2018vandal,rattle,octopus,contro2021ethersolve,pasqua2023enhancing}.
As mentioned in \S\ref{subsec:EVM}, the jump operands are obtained from the stack.
For some BBs, their successors can only be achieved with contexts in the stack, necessitating stack emulation.
Stack emulation~\footnote{Stack emulation represents various terms from prior works, such as stack model~\cite{zhou2018erays}, symbolic stack execution~\cite{contro2021ethersolve,pasqua2023enhancing}, and symbolic execution~\cite{brent2018vandal}.} symbolically emulates the state of the stack during traversal.
For each instruction, instead of real execution, only their effects on the stack (e.g., popping and pushing~\cite{wood2014ethereum}) are emulated.
Some works~\cite{contro2021ethersolve,pasqua2023enhancing} only keep symbols for jump instructions in the stack, while others~\cite{rattle,brent2018vandal} keep symbols for all instructions.
Our work takes the latter method.

\begin{figure*}[t]
    \centering
    \subfigure[Example code]{
        \centering
        \includegraphics[width=0.11\linewidth]{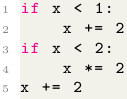}
        \label{fig:motivating_code}
    }
    \subfigure[Reuse-sensitive CFG]{
        \centering
        \includegraphics[width=0.19\linewidth]{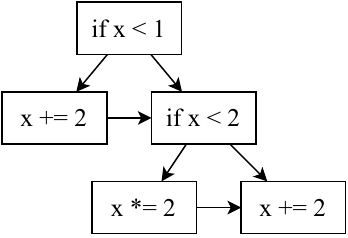}
        \label{fig:motivating_cfg_right}
    }
    \subfigure[Reuse-insensitive CFG]{
        \centering
        \includegraphics[width=0.155\linewidth]{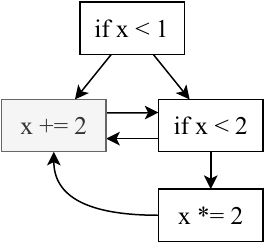}
        \label{fig:motivating_cfg_wrong}
    }
    \subfigure[Reuse-sensitive data analysis]{
        \centering
        \includegraphics[width=0.22\linewidth]{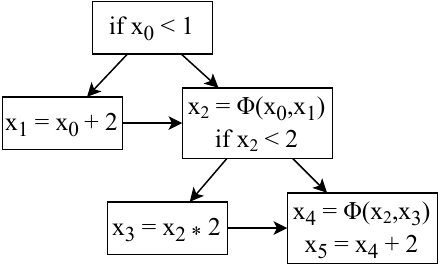}
        \label{fig:motivating_ssa_right}
    }
    \subfigure[Reuse-insensitive data analysis]{
        \centering
        \includegraphics[width=0.21\linewidth]{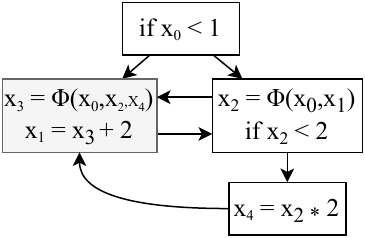}
        \label{fig:motivating_ssa_wrong}
    }
    \vspace{-0.3cm}
    \caption{Reuse-sensitive and reuse-insensitive CFGs and data analysis for the example. The reused BB is in gray background.}
    \vspace{-0.5cm}
\end{figure*}


\section{Motivating Examples}
\label{sec:motivating}


Code reuse is common in EVM bytecode, primarily motivated by the need to economize transaction fees in deployment, with a capped bytecode size limit of 24,576 bytes.
Deploying smart contracts incurs transaction fees that escalate in tandem with contract size.
Consequently, compilers tend to compress contract size by reusing identical code snippets~\cite{grech2022elipmoc,compiler_optimize}.

In this section, we use an example to illustrate the notion of code reuse in EVM bytecode and emphasize the necessity of reuse-sensitive CFGs for static analysis.
For better understanding, we present the example using source code, while acknowledging that our work focuses on EVM bytecode.
The example code is shown in Fig.~\ref{fig:motivating_code}, including repetitions of the statement (\code{x+=2}) at both line 2 and line 5.
As a result, the EVM bytecode that reuses the code (\code{x+=2}) will be generated.

Based on the EVM bytecode with reused code, the constructed reuse-sensitive and reuse-insensitive CFGs are shown in Fig.~\ref{fig:motivating_cfg_right} and Fig.~\ref{fig:motivating_cfg_wrong}, respectively.
In the reuse-sensitive CFG, the control flow and semantics mirror those of the source code, with two conditional jumps leading to four possible execution paths.
However, a discrepancy arises in the reuse-insensitive CFG, where line 1, 3, and 4 all direct to the code \code{x+=2}, making two unintended loops that do not exist in the source code.
The semantics in the reuse-insensitive CFG changes into a dead loop infinitely increasing the value of \code{x}.

Compared to the reuse-sensitive CFG, the reuse-insensitive CFG introduces redundant control-flow dependencies.
In the reuse-sensitive CFG and the source code, line 3 exclusively relies on lines 1 and 2 for control flow.
However, in the reuse-insensitive CFG, line 3 depends on all other three BBs, introducing a false positive.
The redundant control-flow dependencies create infeasible paths that can mislead subsequent analyses such as path constraints.
For example, in the reuse-insensitive CFG, \code{x+=2} can execute under any condition, unlike in practice where it only executes if \code{x} is less than 1.

The redundant control-flow dependencies also cause false positives in data-flow dependency.
We use Static Single Assignment (SSA) form, a prevalent primitive, to illustrate data flows.
Generating SSA form~\cite{goos_simple_2000,cytron_efficiently_1991,rosen_global_1988} involves traversing paths in CFGs and assigning new symbols upon variable definition.
In cases of multiple values at join nodes, $\phi$ instructions are created to represent alternatives.
Fig.~\ref{fig:motivating_ssa_right} and Fig.~\ref{fig:motivating_ssa_wrong} depict the data flows based on reuse-sensitive and reuse-insensitive CFGs, respectively.
In the reuse-sensitive SSA form, $x_2$ depends on $x_0$ and $x_1$ , aligning with the source code where \code{x} in line 3 has a data-flow dependency on \code{x} from lines 1 and 2.
However, in the reuse-insensitive SSA form, $x_2$ depends on all variables, introducing three false positives.

\begin{figure}[t]
    \centering
    \subfigure[The regular jump]{
        \centering
        \includegraphics[width=0.3\linewidth]{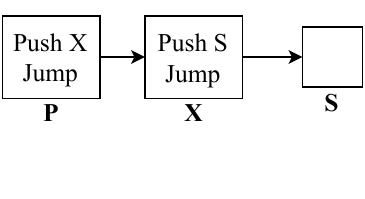}
        \label{fig:noreuse_jump}
    }
    \hspace{0.05\linewidth}
    \subfigure[The jump for code reuse]{
        \centering
        \includegraphics[width=0.35\linewidth]{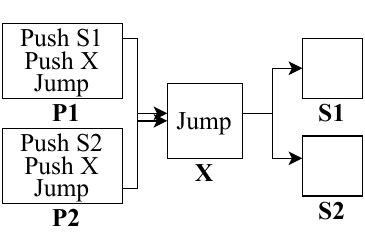}
        \label{fig:reuse_jump}
    }
    \vspace{-0.3cm}
    \caption{Two different jump patterns in EVM bytecode.}
    \label{fig:jump_patterns}
    \vspace{-0.5cm}
\end{figure}

\section{Observations about Code Reuse}
\label{sec:BB_reuse}



Based on our analysis of EVM bytecode and the compiler~\cite{solc_jump_patterns,solc_reuse_jump}, we find that different jump patterns~\cite{contro2021ethersolve,pasqua2023enhancing,grech2022elipmoc,chen2019large} facilitate code reuse.
Fig.~\ref{fig:jump_patterns} depicts the two most common jump patterns.
As shown in Fig.~\ref{fig:noreuse_jump}, a regular jump involves pushing the jump target immediately before the jump instruction, following the sequence "\code{PUSH offset}, \code{JUMP/JUMPI}".
Otherwise, Fig.~\ref{fig:reuse_jump} shows the jump pattern for reuse, where BB \code{X} is reused.
The jump target of \code{X} is not pushed inside itself, but in its predecessors \code{P1} and \code{P2} instead, enabling \code{X} to be reused by two unrelated paths (i.e., \code{P1$\to$S1} and \code{P2$\to$S2}) at runtime.

Based on different jump patterns, code reuse manifests in various forms.
We abstract eight reuse patterns from numerous contracts, classified into two categories: \textit{reuse without control-flow structures} and \textit{reuse with control-flow structures}, posing distinct challenges for static analysis.
These patterns will be discussed in detail in the following.
Notable is that code reuse in practice is more complex, e.g., various patterns may interact with one another in intricate ways.



\begin{figure}[t]
    \centering
    \subfigure[Basic fake join node]{
        \centering
        \includegraphics[width=0.4\linewidth]{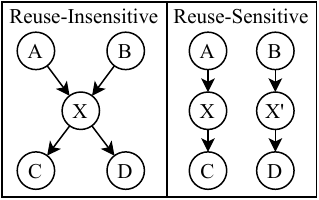}
        \label{fig:reuse_merge_0}
    }
    \subfigure[Basic fake loop]{
        \centering
        \includegraphics[width=0.4\linewidth]{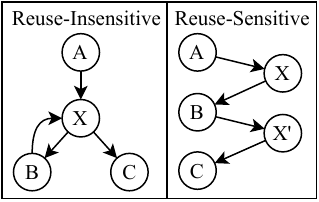}
        \label{fig:reuse_loop_0}
    }
    \subfigure[Sequences of fake join nodes]{
        \centering
        \includegraphics[width=0.4\linewidth]{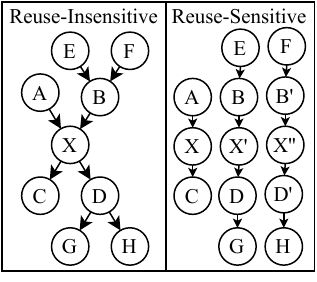}
        \label{fig:reuse_merge_2}
    }
    \subfigure[Nested fake loops]{
        \centering
        \includegraphics[width=0.4\linewidth]{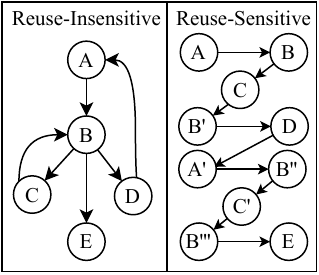}
        \label{fig:reuse_loop_2}
    }
    \vspace{-0.3cm}
    \caption{Reuse patterns without real control-flow structures.}
    \label{fig:reuse_merge}
    \vspace{-0.5cm}
\end{figure}

\subsection{Reuse without Control-flow Structures}

First, we introduce four reuse patterns devoid of control-flow structures as shown in Fig.~\ref{fig:reuse_merge}.
These patterns pose a challenge: \textit{identification of reused code}.
It is challenging to statically distinguish reuse patterns and conventional control-flow structures based solely on graph structures, because there exist legitimate cases for each pattern that exhibit identical graph structures but vary in semantics.

\noindent\textbf{Basic fake join node}\tab
The fake join node is the most common and basic reuse pattern, as shown in Fig.~\ref{fig:reuse_merge_0}.
The paths \code{A$\to$X$\to$C} and \code{B$\to$X$\to$D} reuse \code{X}, constructing a fake join node (\code{A$\to$X} and \code{B$\to$X}) and a fake branch node (\code{X$\to$C} and \code{X$\to$D}) in the reuse-insensitive CFG.
Meanwhile, the reuse-sensitive CFG is two split paths \code{A$\to$X$\to$C} and \code{B$\to$X'$\to$D} by generating \code{X'} to avoid overlap.
Due to the reuse of \code{X}, the reuse-insensitive CFG contains two infeasible paths (\code{A$\to$X$\to$D} and \code{B$\to$X$\to$C}).

\noindent\textbf{Basic fake loop}\tab
When there exists data dependency between the paths reusing the same code, a fake loop is formed, as shown in Fig.~\ref{fig:reuse_loop_0}.
The predecessor and the successor of \code{B} both are \code{X}. 
When \code{X} is executed for the first time, its successor is \code{B}, and for the second time, the successor is \code{C}.
The reuse-insensitive CFG contains a loop between \code{B} and \code{X}, and \code{X} has polymorphic jump targets (\code{B} and \code{C}).
However, in reuse-sensitive CFG, there is only one path.
The fake loop in the reuse-insensitive CFG can mislead following analysis.

\noindent\textbf{Sequences of fake join nodes}\tab
Fig.~\ref{fig:reuse_merge_2} shows a sequence of fake join nodes led by nested code reuse.
The paths \code{A$\to$X$\to$C} and \code{E$\to$B$\to$X$\to$D$\to$G} share \code{X}, while the path \code{F$\to$B$\to$X$\to$D$\to$H} reuses \code{B}, \code{X}, and \code{D}.
The nested reuse makes a sequence of fake join nodes (\code{B} and \code{X}) and fake branch nodes (\code{X} and \code{D}) in the reuse-insensitive CFG.
For blocks \code{C}, \code{G}, and \code{H}, they all have two redundant predecessors from the other two paths.
As a result, the reuse-insensitive CFG contains nine paths while the reuse-sensitive CFG only contains three.
These false positives in reuse-insensitive CFG increase exponentially when fake join nodes appear more times.
This pattern poses a fact that one BB can be reused by different paths with different scope.
The block \code{X} in the paths \code{A$\to$X$\to$C} and \code{E$\to$B$\to$X$\to$D$\to$G} is reused as a single BB, while in paths \code{E$\to$B$\to$X$\to$D$\to$G} and \code{F$\to$B$\to$X$\to$D$\to$H} is reused as a sequence of BBs (\code{B$\to$X$\to$D}).
This fact indicates that \system needs to identify the reused code based on the executed path to build reuse-sensitive CFGs.



\noindent\textbf{Nested fake loops}\tab
Similar to fake join nodes, fake loops can also be nested, as shown in Fig.~\ref{fig:reuse_loop_2}.
There are two snippets of reused code with different scopes.
The first only contains the block \code{B}, while the second contains blocks \code{A}, \code{B}, and \code{C}.
The second includes the first.
The second actually reuses the block \code{B} twice, and the block \code{B} is executed four times.
These fake loops require dual processing to construct the reuse-sensitive CFG, necessitating iterative handling rather than a singular cloning process, as the inner loop impacts the outer loop.

\begin{figure}[t]
    \centering
    \subfigure[Fake join node with a real one]{
        \centering
        \includegraphics[width=0.41\linewidth]{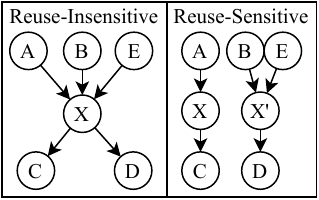}
        \label{fig:reuse_merge_1}
    }
    \subfigure[Fake loop with a real loop]{
        \centering
        \includegraphics[width=0.4\linewidth]{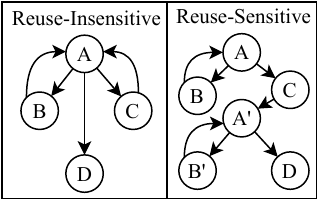}
        \label{fig:reuse_loop_1}
    }
    \subfigure[Fake join nodes with multiple exits]{
        \centering
        \includegraphics[width=0.4\linewidth]{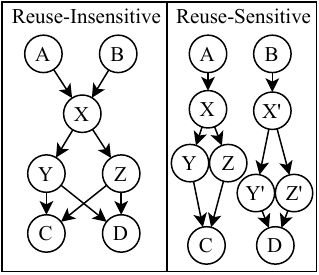}
        \label{fig:reuse_merge_mul_exits}
    }
    \subfigure[Fake loop with control-flow transfers]{
        \centering
        \includegraphics[width=0.4\linewidth]{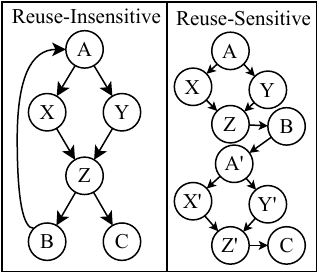}
        \label{fig:reuse_loop_with_merge}
    }
    \vspace{-0.3cm}
    \caption{Reuse patterns with real control-flow structures.}
    \label{fig:reuse_complex}
    \vspace{-0.5cm}
\end{figure}

\subsection{Reuse with Control-flow Structures}

Second, we introduce the reuse patterns that involve real control-flow structures as shown in Fig.~\ref{fig:reuse_complex}.
Code reuse mixed with real control-flow structures is more complex.
These patterns present a new challenge that is \textit{distinguishing and maintaining the real control-flow transfers in the reused code}, to ensure the correctness of the semantics in generated CFGs.

\noindent\textbf{Fake join node with a real one}\tab
When the fake join node is also a real join node, a mixed fake join node is formed, as shown in Fig.~\ref{fig:reuse_merge_1}.
The paths \code{B$\to$X$\to$D} and \code{E$\to$X$\to$D} join at \code{X} while the path \code{A$\to$X$\to$C} reuses \code{X}, forming a mixed fake join node.
The control flow into \code{X} is mixed up with a real join (\code{B$\to$X} and \code{E$\to$X}) and a fake join (\code{A$\to$X}).
The constructed reuse-insensitive CFG looks like a join node with 3 in degrees and 2 out degrees.
This CFG indicates that three predecessors (\code{A}, \code{B}, and \code{E}) merge at \code{X} and then the control flow splits to two successors (\code{C} and \code{D}).
The following analysis based on this CFG will regard both \code{C} and \code{D} have control-flow dependency on all \code{A}, \code{B}, and \code{E}, which conflicts with the real semantics.
To provide accurate control-flow dependency for this pattern, \system needs to recognize the fake join from the real join nodes, which poses a new challenge.
Besides identifying and cloning the reused block \code{X}, the connection between predecessors and successors should be recovered to avoid duplicate copies of reused code or the generation of another fake join node.
\noindent\textbf{Fake join nodes with multiple exits}\tab
The fake join node can also be a real branch node, leading to multiple exits in reused code, as shown in Fig.~\ref{fig:reuse_merge_mul_exits}.
The reused code is a cluster of BBs, including blocks \code{X}, \code{Y}, and \code{Z}.
Blocks \code{Y} and \code{Z} both are the ends of the code reuse.
Block \code{X} has a conditional jump to \code{Y} and \code{Z} (real branch node), and \code{Y} and \code{Z} both jump to the same successor (real join node).
The reuse-insensitive CFG indicates \code{Y} and \code{Z} both have polymorphic jump targets \code{C} and \code{D}, which is inconsistent with the real semantics.
In fact, it is impossible that \code{Y} goes to \code{C} while \code{Z} goes to \code{D} at the same time.


\noindent\textbf{Fake loop with control-flow structures}\tab
Similar to the fake join nodes, the fake loop can also contain real control-flow structures in the loop body, as shown in Fig.~\ref{fig:reuse_loop_1} and Fig.~\ref{fig:reuse_loop_with_merge}.
In Fig.~\ref{fig:reuse_loop_1}, the fake loop contains a real loop in the loop body.
The real loop consisting of \code{A} and \code{B} is executed twice and reused.
Block \code{A} has a conditional jump whose fall-through edge is to \code{B}.
Jump target of \code{A} is \code{C} for the first time, and \code{D} for the second time.
Due to code reuse, block \code{C} jumps back to \code{A} in the reuse-insensitive CFG, forming a fake loop.
In Fig.~\ref{fig:reuse_loop_with_merge}, the reused code is a cluster of BBs including blocks \code{A}, \code{X}, \code{Y}, and \code{Z}, containing a branch node and a join node.
Block \code{Z} is a real join node as the successor of \code{X} and \code{Y}.
This cluster is reused and executed twice with two different successors \code{B} and \code{C}.
In the reuse-insensitive CFG, the reuse constructs a fake loop.
These two patterns both require \system to identify the real control-flow structures in the fake loop body.

\begin{figure}[t]
    \centering
    \includegraphics[width = \linewidth]{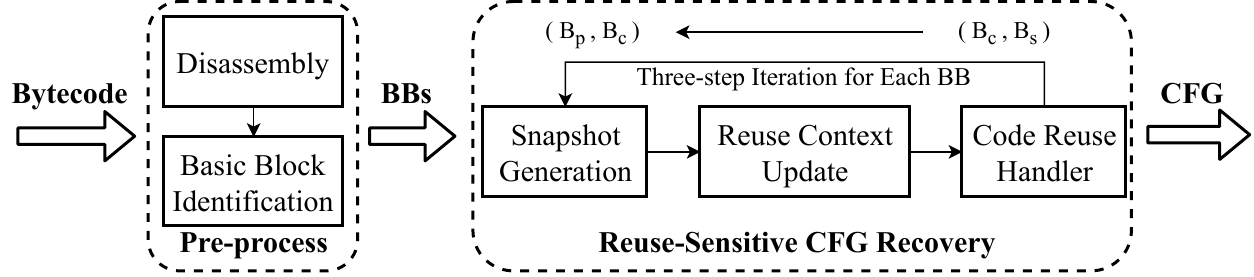}
    \vspace{-0.3cm}
    \caption{High-level architecture of \system.}
    \label{fig:overview}
    \vspace{-0.4cm}
\end{figure}

\begin{table}[t]
    \centering
    \caption{Reference Symbols}
    \label{tab:symbol}
    \vspace{-0.3cm}
    \footnotesize
    \resizebox{\linewidth}{!}{%
    \begin{tabular}{r|l}
    \toprule
    $B_c$ & The BB processed in current iteration \\ 
    $B_p$ & A predecessor of $B_c$, which has been processed in previous iterations \\ 
    $B_s$ & A successor of $B_c$, which will be processed in following iterations \\ 
    $S_{start}$ & The state of stack before emulation \\ 
    $S_{end}$ & The state of stack after emulation \\ 
    \bottomrule
    \end{tabular}
    \vspace{-0.4cm}
    }
\end{table}


\section{Design}
\label{sec:design}

In this section, we present \system which identifies code reuse and constructs reuse-sensitive CFGs for EVM bytecode.
To enhance comprehension, we offer illustrative examples in \S\ref{sec:design_example}.
We introduce the core idea of \system at first.
\S\ref{sec:BB_reuse} proposes two challenges due to code reuse: \textit{identification of code reuse} and \textit{maintaining real control-flow transfers}, with various reuse patterns.
To address these two challenges and handle all patterns, \system adopts a dynamic approach to identify code reuse during CFG construction.
Fundamentally, \system defines \textit{reuse context} to identify code reuse generally.

\noindent \textbf{Definition 1} \tab For each BB, its reuse context is the jump operands that are pushed by predecessors in its stack, which are also called pre-pushed jump operands in the following.

The definition of reuse context is based on our insights in \S\ref{sec:BB_reuse}. First, if the jump operand of a BB is pre-pushed, it indicates that this BB can be reused, which can be used to identify code reuse. Second, the values of pre-pushed jump operands from different predecessors are the same in real join nodes, and are different in fake join nodes, which can be used to maintain real control-flow transfers.
Therefore, pre-pushed jump operands can be reuse contexts.
If the contexts from two predecessors are the same, these two predecessors can be connected to the same successor.
Otherwise, it is code reuse, and a new clone of the successor should be generated.

Based on this core idea, the overview of \system is shown in Fig.~\ref{fig:overview}, and the reference symbols are shown in Table~\ref{tab:symbol}.
\system consists of two components: pre-process and reuse-sensitive CFG recovery.
Pre-process takes bytecode as input and outputs identified BBs as nodes in the CFG.
Reuse-sensitive CFG recovery connects control-flow transfers between BBs, and identifies code reuse during the process.
To construct CFGs, \system adopts an iterated method to recursively visit BBs.
In the recursive traversal, each BB can be visited multiple times, and \system needs to identify code reuse since the second visit by comparing reuse contexts cross visits.
Therefore, \system keeps the snapshots about stack state of each BB for every visit (\textit{snapshot generation}), and taints reuse context (\textit{reuse context update}) in every iteration, to ensure all connections in the generated CFG are non-reuse (\textit{code reuse handler}).
The iterations start from the BB at offset 0x0, and end when all control-flow transfers are found.
Each iteration emulates $B_c$ with context from $B_p$, and connects $B_s$ to $B_c$.
The $B_c$ becomes $B_p$ for the next iteration, and $B_s$ becomes $B_c$.

\subsection{Pre-process}
\label{subsec:pre-process}

Pre-process consists of disassembly and basic block identification.
Disassembly translates the hex string into readable mnemonics with offsets, and basic block identification splits monolithic mnemonics into BBs which will be nodes in CFGs.

\noindent\textbf{Disassembly}\tab
\system uses linear sweep strategy in disassembly, to scan bytecode, convert each instruction from hex string into its corresponding mnemonic, and increase the program counter with the length of the instruction.
The mnemonic and length of each instruction are recorded in the formal specification of Ethereum~\cite{wood2014ethereum}.
In addition, \system skips embedded data of \code{PUSH} instructions to avoid unintended instructions.
If data of \code{PUSH} instructions are converted into instructions, the disassembled instructions will conflict with the real execution.

\noindent\textbf{Basic block identification}\tab
\system identifies BBs by locating \code{JUMPDEST} instruction which is used to mark the start of a BB.
Unlike previous work~\cite{contro2021ethersolve,pasqua2023enhancing,rattle,octopus} that directly strips data before identifying BBs, \system keeps all data in this process to ensure the completeness of generated CFGs.
That is because developers can deliberately embed code in the data region for the purpose of code obfuscation.
For instance, the code can be put inside the data region of the \code{CODECOPY} instruction or the metadata section.
Removing data and metadata too early can lead to an incomplete CFG.



\subsection{Reuse-sensitive CFG Recovery}
\label{subsec:construct_cfg}

\SetKwComment{Comment}{\ensuremath{\triangleright} }{}

\begin{algorithm}[t]
    \scriptsize
  \SetCommentSty{textbf}
  
  \SetKwFunction{PrepareStack}{PrepareStack}
  \SetKwFunction{Emulate}{Emulate}
  \SetKwFunction{Snapshot}{Snapshot}
  \SetKwFunction{UpdateContext}{UpdateContext}
  \SetKwFunction{ReuseHandler}{ReuseHandler}
  
  \caption{Reuse-sensitive CFG Recovery}
  \label{alg:overflow}
  \KwIn{$BBs$ \Comment*{All BBs identified in pre-process} }
  \KwOut{$CFG$ }
  $B_c \leftarrow $ The BB at offset 0x0  \Comment*{Initialization}
  $work\_list \leftarrow [(None, B_c)]$\;
  \While{$work\_list$ is not empty}{
    $B_p, B_c \leftarrow work\_list.pop()$ \;
    \If{$B_p$ is None}{$B_c.S_{start} \leftarrow $An empty stack\;}
    \Else{
    $B_c.S_{start}\leftarrow $ \PrepareStack{$B_p$,$B_c$}\;}
    $B_c.S_{end}, \mathit{Off}_s \leftarrow$ \Emulate{$B_c$}
    \SetCommentSty{textsf}
    \tcc{$\mathit{Off}_s$ is a list containing offsets of all $B_c$'s successors}
    \SetCommentSty{textbf}
    \Snapshot{$B_c.S_{start}$,$B_c.S_{end}$} \Comment*{\S\ref{subsubsec:snapshot_generation}}
    \For{$\mathit{offset}$ in $\mathit{Off}_s$}{
        \If{$\mathit{offset}$ in $B_c.S_{start}$}{\UpdateContext{$\mathit{offset}$} \Comment*{\S\ref{subsubsec:update_context}}}
        $B_s \leftarrow$ \ReuseHandler{$B_c,\mathit{offset}$} \Comment*{\S\ref{subsubsec:reuse_handler}}
        $CFG.add\_edge(B_c, B_s)$\;
        $work\_list.push(B_c, B_s)$\;
        }
    }
\end{algorithm}


The reuse-sensitive CFG recovery is an iterated and incremental process, as shown in Algorithm~\ref{alg:overflow}.
The whole process starts from the first BB at code offset 0x0 (Line 1), with an empty stack as its $S_{start}$ (Line 6), and stops when no new control-flow transfers are found.
In each iteration, a BB is symbolically emulated with the stack inherited from its predecessors, and its successors are obtained for following iterations.
There are three steps involved in each iteration.

\noindent \textbf{Step I: Snapshot generation.}
The first step works towards $B_c$, including stack emulation and snapshot preservation (Line 4-12).
\system symbolically emulates $B_c$, and keeps $S_{start}$ and $S_{end}$ as the snapshot of $B_c$ in this visit.
This snapshot will be used to compare with other snapshots in future visits to decide whether $B_c$ is reused (in step III).

\noindent \textbf{Step II: Reuse context update.}
This step works towards $B_p$s and their clones, updating their reuse context based on taint analysis.
The reuse context is pre-pushed jump operands in the $S_{start}$, which are tainted when the operands are used.
In this step, \system uses the jump operand used in $B_c$ as the initial source to collect taint sources and transfer taints.

\noindent \textbf{Step III: Code reuse handler.}
The third step works towards $B_s$s, detecting reuse of them based on the snapshots with tainted reuse context generated in Step I and Step II from previous visits.
In this step, \system selects a non-reuse $B_s$ as jump target of $B_c$, or generates a new clone as $B_s$ if needed.

After that, the iteration for $B_c$ is finished. And \system continues to the next iteration for $B_s$ (which will be $B_c$ in that iteration).
Each step in the iteration affects different BBs instead of just $B_c$.
Therefore, iterations can interact with each other, helping to construct a reuse-sensitive CFG.
In the following, we will elaborate on each step.



\subsubsection{Snapshot Generation}
\label{subsubsec:snapshot_generation}

The initial step in each iteration involves creating snapshots for $B_c$. These snapshots encompass $S_{start}$ and $S_{end}$, capturing the stack state before and after emulation, respectively. The generation of $S_{start}$ occurs in Algorithm~\ref{alg:overflow} Lines 5-10. For the initial BB with an offset of 0x0, its $S_{start}$ remains an empty stack. In contrast, for other BBs, their $S_{start}$ is inherited from $B_p.S_{end}$ (Line 9). In cases where $B_c$ has additional predecessors with differing $S_{end}$ values from $B_p.S_{end}$, $\phi$ instructions are introduced to represent multiple potential values. These $\phi$ instructions facilitate the replacement of varying stack values to establish the new $B_c.S_{start}$.

Following the acquisition of $S_{start}$ for $B_c$, the stack emulation process proceeds to derive $S_{end}$ using $S_{start}$ as the execution context (Line 11). As elucidated in \S\ref{subsec:static_analysis}, \system engages in stack emulation and leverages symbols in SSA format to denote the outcomes of emulated instructions. Each instruction involves the popping of stack elements as instruction operands, with the subsequent generation of an SSA symbol for pushing back to the stack as the return value if the instruction yields one. Post-emulation, all instructions within $B_c$ are translated into three-address codes annotated with their operands and return values represented by SSA symbols or constants.


\subsubsection{Reuse Context Update}
\label{subsubsec:update_context}
Reuse context is identified based on taint analysis, including taint generation and taint transfer.
These taints are used in identification of code reuse (Step III).

\noindent\textbf{Taint generation}\tab
After stack emulation, the jump target of $B_c$ is determined.
If this target has been pre-pushed, \system utilizes it as the initial source and generates taint for it, indicating that it belongs to reuse context.
That is because different pre-pushed jump targets align with the code reuse pattern, signifying reused code (\S\ref{sec:BB_reuse}).

Based on the initial source, \system collects others taint sources by backwards data analysis.
By utilizing SSA symbols generated in stack emulation (Step I), \system builds def-use chains for the initial source, and traces back to the origin, i.e., the \code{PUSH} instruction that introduces the operand into the stack.
During this process, \system taints all intermediate variables as sources.
For example, if the variable \code{a} is the initial source and is calculated by \code{AND(0xffff,b)}, \code{b} is also tainted as a source.

\noindent\textbf{Taint Transfer}\tab
Reuse context taints propagate from sources based on data dependencies. Moreover, taints are also transferred among clones of the same BB, as these clones can share locations of reuse context with each other. This sharing arises from the fact that identical instructions yield the same effects on the stack, involving an equal number of stack pops and pushes. Consequently, pre-pushed jump operands that should be tainted as reuse context are stored in the same identical locations in these clones.
For a pair of clones, same values of pre-pushed jump operands signify the execution of the same instructions, leading to identical locations for the subsequent pre-pushed jump operand. Conversely, different pre-pushed jump operands between two clones, denoting distinct paths that execute different instructions, signal the termination of shared reuse context.

Based on these insights, taint transfers between clones following these steps.
Initially, \system aggregates tainted reuse contexts from clones of a basic block, denoted by \code{(stack index, value)}. Here, the \code{stack index} signifies the position of the tainted reuse context in the stack, while the \code{value} represents the value of the tainted variable. Subsequently, \system facilitates the transfer of taints between clones that possess identical values at corresponding locations.

\subsubsection{Code Reuse Handler}
\label{subsubsec:reuse_handler}

The third step is to detect code reuse when selecting $B_s$ for $B_c$ and generate clones if necessary.
To build a reuse-sensitive CFG, \system needs to select a BB that is not reused as $B_s$ and connect it to $B_c$.
Snapshots of $B_s$ in different iterations are utilized to ensure the connection will not lead to code reuse.
Specifically, \system can obtain the offset of $B_s$ from emulated jump instructions in $B_c$ (Line 11 in Algorithm~\ref{alg:overflow}).
Because of cloning in previous iterations, the same offset can lead to multiple BBs.
All these BBs are candidates for successors of $B_c$.
\system compares the $B_c.S_{end}$ with all candidates to select a non-reuse successor as $B_s$.
If there are no suitable BBs, a new clone will be generated.


The selection of the successor is based on the snapshots generated by the Step I (\S\ref{subsubsec:snapshot_generation}) with reuse context marked by the Step II (\S\ref{subsubsec:update_context}) in previous iterations.
Specifically, \system compares $B_c.S_{end}$ with $S_{start}$ in snapshots of each candidate.
Given the two stacks, \system regards it is not code reuse iff all the values of tainted reuse contexts are the same.
If \system cannot find a non-reuse candidate for $B_c$, a new clone will be generated as $B_s$ in the end.
When generating the new clone, context sharing strategy (\S\ref{subsubsec:update_context}) is also adopted to initialize the reuse context of the new clone, to ensure the reuse context of the new clone is sufficient for further reuse identification in following iterations.


\subsection{Example}
\label{sec:design_example}

\begin{figure*}[t]
    \centerline{\includegraphics[width = 0.97\linewidth]{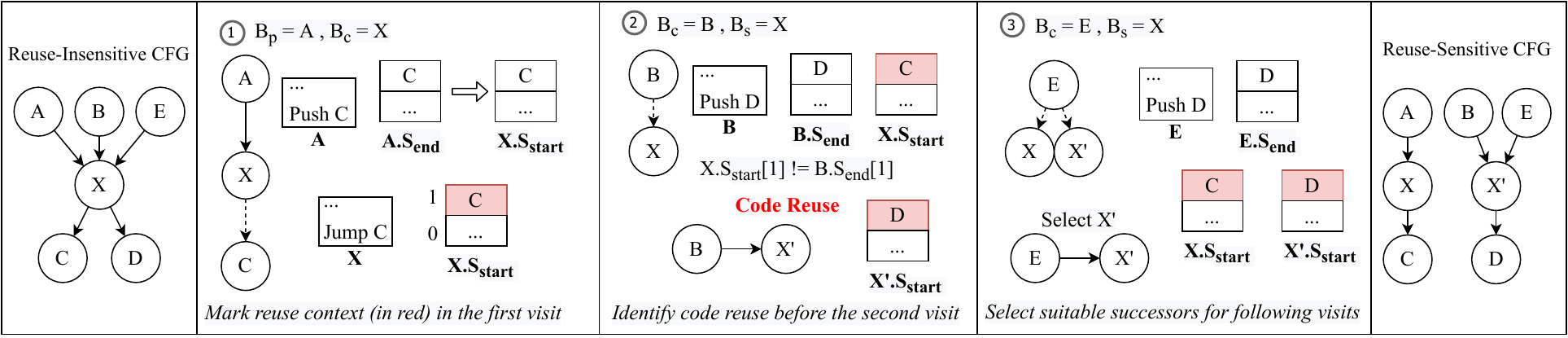}}
    \vspace{-0.2cm}
    \caption{An example to illustrate the basic idea of \system.}
    \label{fig:design_example_basic}
    \vspace{-0.5cm}
\end{figure*}

An example is used to illustrate the previous steps about code reuse identification (\S\ref{subsec:construct_cfg}), as shown in Fig.~\ref{fig:design_example_basic}.
\footnote{We also provide another example in our github repo which is more complex to explain taint transfers in marking reuse context (\S\ref{subsubsec:update_context})}.
Note that, examples only show snippets instead of the whole programs.
We use $BB_\mathit{off}$ to denote offset of BBs.
In the example, Block \code{X} is a mixed fake join node with a real one.
There are three steps to construct a reuse-sensitive CFG for it.

\noindent\textbf{Step 1: Mark reuse context for \code{X}.} For the first visit, \code{X} inherits $S_{start}$ from \code{A} whose $S_{end}$ contains $C_\mathit{off}$.
After emulation, \system finds that the used jump operand $C_\mathit{off}$ is pre-pushed and exists in $X.S_{start}$.
Therefore, \system taints the $C_\mathit{off}$ in $X.S_{start}$ as \code{X}'s reuse context.
Then \system explores the path recursively until the end.

\noindent\textbf{Step 2: Identify code reuse with the marked reuse context.}
When exploring other paths, \system emulates \code{B} and needs to find the successor for it.
\system collects all candidates, which are \code{X}.
\system compares the values of tainted reuse context in $X.S_{start}$ and $B.S_{end}$.
Then \system finds that they are not equal as one is $C_\mathit{off}$ while another is $D_\mathit{off}$, which means that \code{A} and \code{B} reuse \code{X}.
As a result, \system cannot find a suitable successor for \code{B}, and generates a new clone \code{X'}.
After generation, \code{X} shares reuse context with \code{X'}.
\system first gathered locations and values of tainted reuse context from \code{X}, which is $C_\mathit{off}$ at index 1.
Then, \system transfers the taint to the same location in \code{X'}, which is $D_\mathit{off}$ at index 1.
After that, \system finds that \code{X} and \code{X'} do not have the same value at this location, and finishes the sharing.

\noindent\textbf{Step 3: Select successors with the marked reuse context.}
\system finds \code{E} whose jump target is also $X_\mathit{off}$.
Now there are two candidates for \code{E}'s successor, i.e., \code{X} and \code{X'}.
\system compares $E.S_{end}$ with $X.S_{start}$ and $X'.S_{start}$, and finds that the contents at reuse context from $E.S_{end}$ and $X'.S_{start}$ are the same, which both are $D_\mathit{off}$.
Therefore, \code{X'} is the successor for \code{E}, and \system connects them.
After that, \system solves the reuse of \code{X} and keeps the real join of \code{B} and \code{E}.

\section{Evaluation}
\label{sec:evaluation}


The evaluation aims to qualify \system by answering the following research questions.

\noindent \textbf{RQ1: Precision.} Do CFGs generated by \system exhibit higher precision by reuse-sensitive analysis? Does the reuse-sensitive analysis influence correctness of CFGs? Can \system successfully identify all code reuse?

\noindent \textbf{RQ2: Performance.} Can \system successfully generate CFGs for real-world smart contracts? What is the performance of \system? Is \system scalable, in terms of analyzing smart contracts in different sizes?

\noindent \textbf{RQ3: Effectiveness.} How effective is \system in solving the downstream static analysis problem?

\begin{figure}[t]
    \centering
    \subfigure[The CDF for contracts size in dataset.]{
        \centering
        \includegraphics[width=0.31\linewidth]{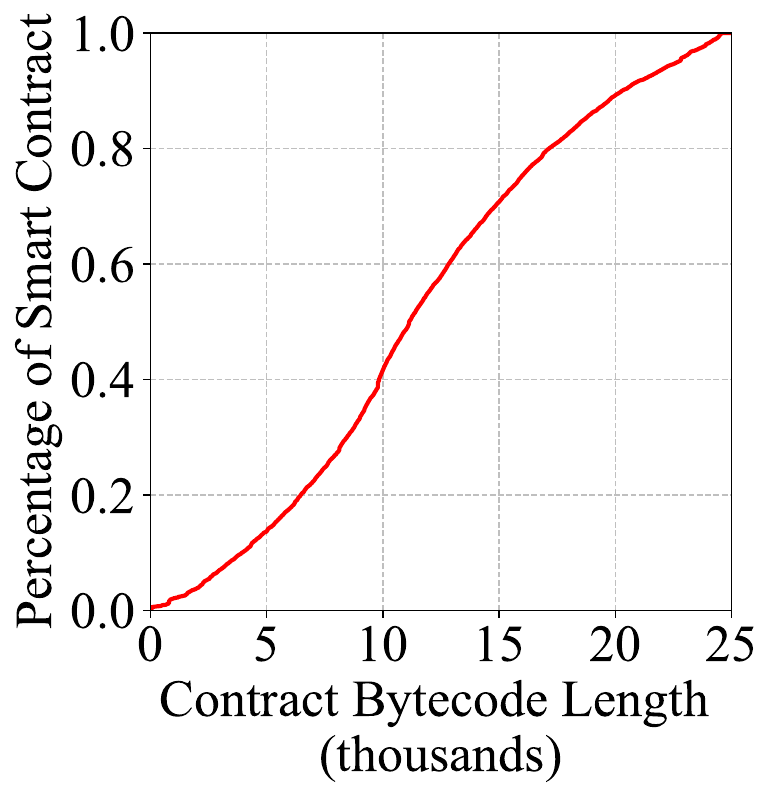}
        \label{fig:dataset_cdf}
    }
    \subfigure[The distribution of compiler versions for contracts in dataset.]{
        \centering
        \includegraphics[width=0.6\linewidth]{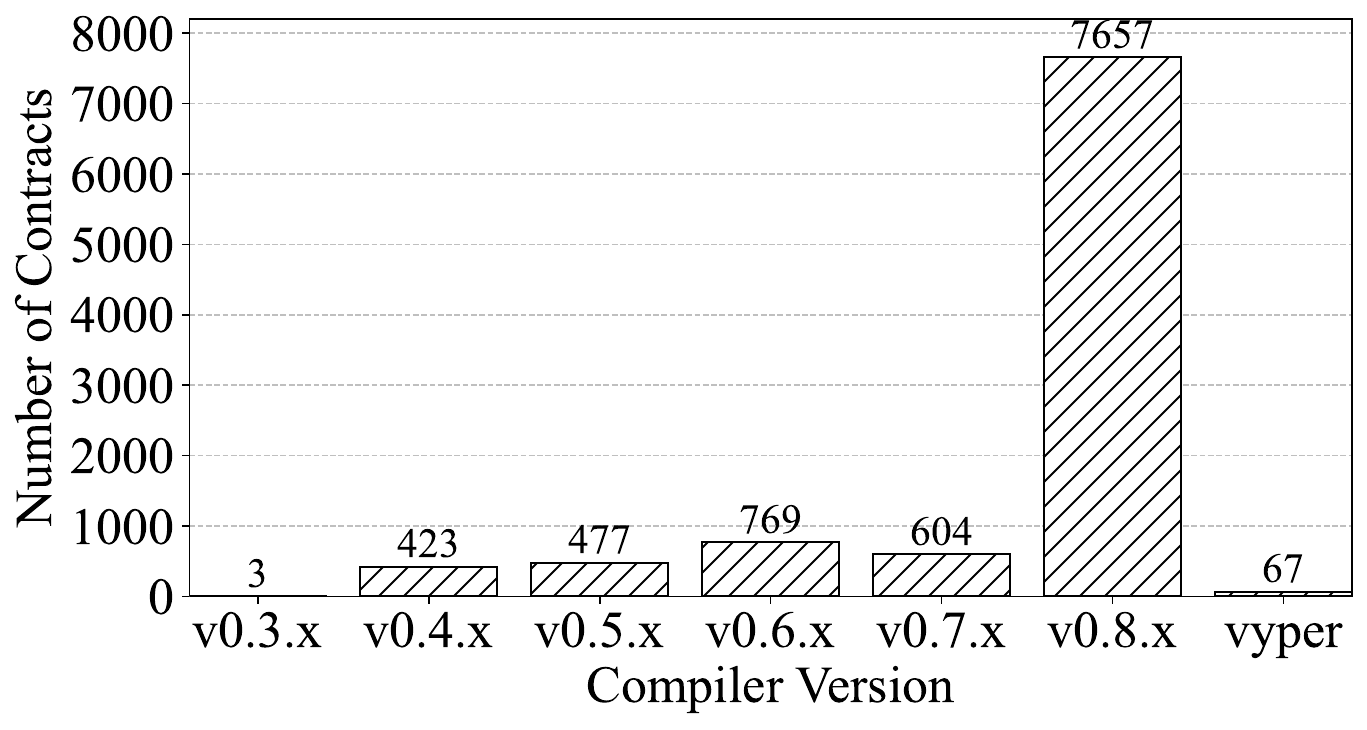}
        \label{fig:dataset_compiler}
    }
    \vspace{-0.3cm}
    \caption{The statistics of the dataset.}
    \vspace{-0.5cm}
\end{figure}

\noindent\textbf{Dataset}\tab
We use two datasets for the evaluation.
For precision and performance, we utilized the most popular smart contracts as our dataset for this evaluation. Specifically, we collected 10,000 unique contracts that had the most transactions from September 27, 2021, to September 27, 2022. We obtained the compile information for these contracts from Etherscan~\cite{etherscan}. The scale and compiler version distribution of the smart contracts in our dataset are shown in Fig.~\ref{fig:dataset_cdf} and Fig.~\ref{fig:dataset_compiler}, respectively.
The size of smart contracts in our dataset varies from tens of bytes to nearly the maximal size limit, covering all possible sizes of smart contracts. This allows us to measure the performance of all tools in handling smart contracts of different sizes, ensuring a comprehensive evaluation of their effectiveness.
The compiler configuration of the smart contracts in our dataset is also diverse. Although most popular smart contracts use the latest version of the Solidity compiler, the oldest version is still used. Additionally, while nearly all popular smart contracts use the Solidity compiler, some contracts use Vyper. This diversity of compiler configurations requires the ability of analyzing EVM bytecode with different versions of different compilers, making our evaluation comprehensive and applicable to a wide range of smart contracts.

For effectiveness, we use SolidiFI dataset~\cite{SolidiFI-benchmark}, which consists in 50 Ethereum smart contracts whose source code has been injected with 9,369 security vulnerabilities by the automated approach SolidiFI~\cite{ghaleb2020effective}.
This dataset has been used to compare the most prominent vulnerability detection tools in previous works~\cite{ghaleb2020effective,contro2021ethersolve,pasqua2023enhancing}.

\noindent\textbf{Comparison Tools}\tab
In this evaluation, we compare \system with six state-of-the-art (SoA) tools, namely Rattle~\cite{rattle}, Vandal~\cite{brent2018vandal}, Ethersolve~\cite{contro2021ethersolve,pasqua2023enhancing}, Gigahorse~\cite{grech2019gigahorse,grech2022elipmoc}, Octopus~\cite{octopus}, and Mythril~\cite{mythril}.
These tools serve for different aims and use different methodologies to generate CFGs.
Rattle and Vandal are two static analysis frameworks, while Ethersolve is a specialized tool that focuses on CFGs.
Gigahorse is one of the leading decompilers of EVM bytecode.
Octopus uses static primitives to detect vulnerabilities in smart contracts, while Mythril analyzes EVM bytecode with symbolic execution and constraint solving techniques.
We find that Octopus crashes due to unsupported operations and fix them because we are more interested in comparing algorithms in CFG generation than the imperfection of implementations


\subsection{Precision}
\label{subsec:precision}

To answer RQ1, we measure the precision of \system from path quality in CFGs and the capability to detect code reuse.

\subsubsection{Path quality}

We measure the quality of paths in a CFG to measure the precision of the CFG, for the following three reasons.
First, paths contain information on control-flow dependencies, which is lossless in representing a CFG.
Second, as mentioned in \S\ref{sec:BB_reuse}, code reuse can result in a geometric increase in the number of possible paths in CFGs.
Therefore, a larger number of possible paths can suggest the presence of reused code in CFGs.
Third, paths can reflect the efficiency of subsequent analyses based on the CFG, indicating the applicability of the tool.

\noindent\textbf{Metric}\tab
We use two metrics to measure the path quality of CFGs: the number of paths in CFGs generated by all tools, and coverage of execution traces extracted from historic transactions.
A higher coverage with fewer paths indicates that the CFG contains more real execution traces while removing infeasible paths introduced by code reuse, which is better in precision of control-flow information.
We do not use real execution traces as ground truth, because real transactions only trigger a small part of contracts as most feasible paths are used for error handling which are not toughed by transactions.


\noindent\textbf{Method}\tab
To measure path number, we take \system as base and use ratio to represent the number of paths in CFGs generated by each tool, as the number of paths can be very large due to path explosion.
To ensure a fair comparison, for each compared tool, we only compare smart contracts for which both \system and the compared tool can successfully generate CFGs.
In addition, we exclude Mythril due to its low success rate (explained in \S\ref{subsec:performance}).
Specifically, to calculate the ratio of a compared tool, we use the average of the ratios for all common contracts of \system and the compared tool.
For each contract, we use dynamic programming to count paths in the CFGs generated by \system and the compared tool, and get their ratio.
For loops in CFGs, we only count once.

To measure coverage, we collected real execution traces from historical transactions of smart contracts in the dataset.
In total, we collect 252,726,569 transactions and extract 950,221 execution traces from transactions.
Most transactions trigger the same execution trace of a contract, which is served as a functionality of the contract.
To calculate the trace coverage of a CFG, we check the existence of execution traces in the CFG.
After that, we used the average trace coverage of all CFGs to represent the trace coverage of a tool.





\begin{figure}
    \centering
    \includegraphics[width = 0.99\linewidth]{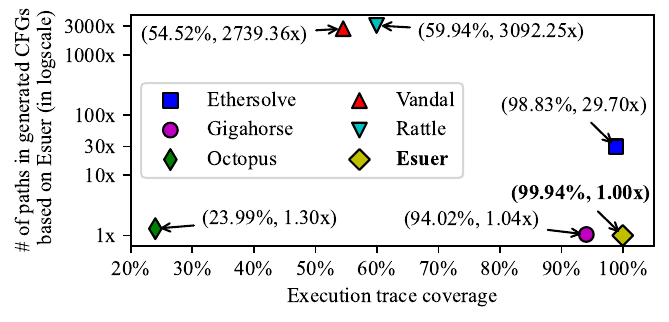}
    \vspace{-0.3cm}
    \caption{The precision of all tools. A higher coverage with fewer paths indicates better precision.}
    \label{fig:precision}
    \vspace{-0.3cm}
\end{figure}


\noindent\textbf{Result}\tab
The results are shown in Fig.~\ref{fig:precision}, demonstrating improvements in path quality achieved by \system's reuse-sensitive analysis.
The y-axis represents the ratio of the number of paths in CFGs generated by each tool to that in \system's CFGs.
The x-axis represents the execution trace coverage of CFGs generated by each tool.
A higher value in x-axis means that the CFGs generated by the tool cover more real execution traces, indicating that CFGs include more feasible paths and are more complete. 
A lower value in y-axis notates generated CFGs contain less paths, indicating there are less infeasible paths in generated CFGs
As a result, \system outperforms all tools.
\system has both the smallest number of paths and the highest execution trace coverage, which means that \system provides the most complete control-flow information with the fewest infeasible paths.
Therefore, it is reasonable to conclude that \system's reuse-sensitive analysis is highly effective in improving the precision of CFGs.


The loss of trace coverage is caused by the absence of edges in CFGs.
For \system, there are 138 execution traces not in the CFGs, involving two smart contracts.
These two contracts both use jump operands stored in the memory.
Similar to most existing static analysis tools towards EVM bytecode , \system do not emulate contents in memory and storage, leading to unknown jump destinations.
The problem is rare in present contracts, as these two contracts are both compiled by old versions of Vyper.

As for other tools, they lose trace coverage due to flaws in their CFG construction method.
There are lots of orphan BBs in CFGs generated by Octopus, which is the reason to both the small amount of paths and low trace coverage.
That is because Octopus misuses magic code to extract code and maintains wrong offset of instructions, leading to failures in connecting BBs.
Vandal and Rattle both recursively connect BBs backward with limited stack depth.
Without reuse analysis, they spend most time and stack depth in infeasible edges due to code reuse, and fail to connect feasible edges that can be executed in practice.
As a result, they both have a great amount of paths in CFGs with low trace coverage.
Gigahorse also uses jump operands as a key feature with pattern matching to recover internal functions and clone BBs that belong to different extracted functions, which reduces the number of paths but also diminishes trace coverage.
Intricate reuse scenarios coupled with the influence of the compilation environment inherently restrict the efficacy of pattern matching in EVM bytecode, particularly in addressing complex corner cases.
Therefore, \system uses a dynamic way to complete reuse contexts and identify code reuse, which is general and accurate regardless of compilation environments.
Ethersolve uses a forward method to construct CFGs, which makes the path amount of it is smaller than Rattle and Vandal.
However, Ethersolve does not handle code reuse, resulting in its path amount nearly 30 times of \system.
In addition, Ethersolve only maintains jump operands in stack, which leads to the loss of edges in CFGs.
In contrast, \system keeps all values in stack, which enables \system find edges that engage calculation.

\subsubsection{Code reuse detection}
\label{subsubsec:reuse_detection}
We measure the capability of code reuse detection of \system in two aspects: correctness of code reuse detection and existence of code reuse in generated CFGs.

\noindent\textbf{Metric}\tab
The correctness of code reuse detection is to check whether the code reuse identified by \system is correct by comparing with the compiler.
When the identified code reuse is also marked as reuse in compiler, it is a true positive (TP).
If the identified code reuse is not marked by compiler, it is a false positive (FP).
If the code reuse is marked by the compiler but not identified by \system, it is a false negative (FN).
We use the precision ($\frac{\left | \mathit{TP} \right |}{\left | \mathit{TP} \right | + \left | \mathit{FP} \right |}$), recall ($\frac{\left | \mathit{TP} \right |}{\left | \mathit{TP} \right | + \left | \mathit{FN} \right |}$), and F1-score ($\frac{2\times Precision \times  Recall}{Precision + Recall}$) metrics to measure the correctness of code reuse detection.
$\|$ denotes the number of TPs/FPs/FNs.

The existence of code reuse in generated CFGs is a measure of the ability of \system to eliminate code reuse.
In particular, we use the existence of polymorphic jump targets to measure the existence of code reuse in CFGs, which is a metric used by previous workse~\cite{grech2019gigahorse,grech2022elipmoc} to measure precision.
Polymorphic jump targets means that jumps resolve to more than one target for non-fallthrough CFG edges, which is abnormal in compiler-generated EVM bytecode.
The root cause of polymorphic jump targets is that multiple predecessors of reused code push different pre-pushed jump operands.
In other assembly languages, there are other cases that can lead to polymorphic jump targets such as jump tables introduced by switch cases or function pointers.
However, in EVM bytecode, these cases are processed by if-else structures, without polymorphic jump targets.
Therefore, polymorphic jump targets can be a characteristic of code reuse in EVM bytecode.


\noindent\textbf{Method}\tab
To get reuse information from the compiler, we extract "tag" information from the assembly.
Assembly has the same semantics as bytecode, and contains more information from the compiler, which is the "tag" information.
Tag is used by the solidity compiler to flag code snippets~\cite{solc_reuse_jump}.
We identify code reuse in the assembly by identifying tag reuse.
Specifically, tag is used in the assembly with the instruction "push tag".
If a tag is used in multiple places, all BBs with the tag are reused.
For example, if there is the instruction "push tag 3" in both tag 1 and tag 2, it means that code in tag 3 is reused by tag 1 and tag 2.
After achieving reused code in the assembly, we compare it with reused code identified by \system.

However, in some cases, the reuse tag in assembly is not actually reused in execution.
For example, two path after a branch both "push tag", but one of them reverts before using the tag.
As a result, the tag is only used once but pushed twice.
Therefore, we further evaluate the existence of code reuse in generated CFGs by "polymorphic jump targets".
Specifically, we count the number of polymorphic jump targets in CFGs.


\noindent\textbf{Result}\tab
\system achieves precision of 98.16\%, recall of 96.18\%, and F1-score of 97.02\%.
The loss of precision is because of FPs, which are introduced by clones of end BBs.
End BBs are the BBs whose terminators end transactions such as STOP and REVERT. 
\system clones end BBs if they have multiple predecessors (explained in \S\ref{sec:discussion}).
However, there do exist some shared exit code in the source code.
As a result, there are some FPs in the generated CFGs.
We think it is acceptable because clones of end BBs do not cause further influences.
In addition, these FPs only lead to a small loss of precision.
The loss of recall is because of FNs, which are introduced by the different mechanisms of \system and assembly to identify code reuse.
As mentioned before, in assembly, reuse is identified by matching the instruction "push tag", but not all the pushed tag are used in real execution.
Meanwhile, \system identifies code reuse by stack emulation, which is more similar to real execution.
Therefore, the tags that are only pushed but not used are not identified by \system, leading to FNs.

However, these FNs do not actually lead to code reuse in generated CFGs.
To prove this, we further count polymorphic jump targets in CFGs generated by \system.
The result is that there are no polymorphic jump targets in \system-generated CFGs.
In the meantime, there are 11,769 polymorphic jump targets in 1,106 CFGs generated by Gigahorse.
The CFGs that contain polymorphic jump targets account for 11.17\% in all Gigahorse-generated CFGs, indicating that Gigahorse does not eliminate code reuse.
The result proves that the method of \system can eliminate code reuse in generated CFGs.


    \vspace{-0.1in}

\begin{framed}
    \vspace{-0.1in}
        \noindent
        \textbf{Answer to RQ1}\tab
        \system achieves the best precision with the fewest paths and highest execution trace coverage of 99.94\% compared with other tools. The code reuse identified by \system achieves a F1-score of 97.02\% in correctness. In addition, there do not exist polymorphic jump targets in CFGs generated by \system, indicating \system successfully eliminates code reuse.
        \vspace{-0.1in}
\end{framed}

\subsection{Performance}
\label{subsec:performance}

\begin{table}[t]
    \centering
    \caption{The performance of all tools.}
    \label{tab:performance}
    \vspace{-0.3cm}
    \footnotesize
    \resizebox{\linewidth}{!}{%
    \begin{tabular}{lccccccc} 
    \toprule
    Performance & Ethersolve & Mythril & Gigahorse & Octopus & Vandal & Rattle & \system \\
    \midrule
    Success Rate & 98.81\% & 5.48\% & 99.04\% & 60.57\% & 94.04\% & 59.87\% & 99.25\% \\
    Error Rate & 1.14\% & 0.04\% & 0 & 35.23\% & 0.35\% & 30.29\% & 0 \\
    Timeout Rate & 0.05\% & 94.48\% & 0.96\% & 4.20\% & 5.61\% & 9.84\% & 0.75\% \\
    Average Time (s) & 0.86 & 42.93 & 7.40 & 15.23 & 24.10 & 22.45 & 1.06 \\
    \bottomrule
    \end{tabular}
    \vspace{-0.3cm}
    }
\end{table}

\begin{figure*}
    \centering
    \includegraphics[width = 0.99\linewidth]{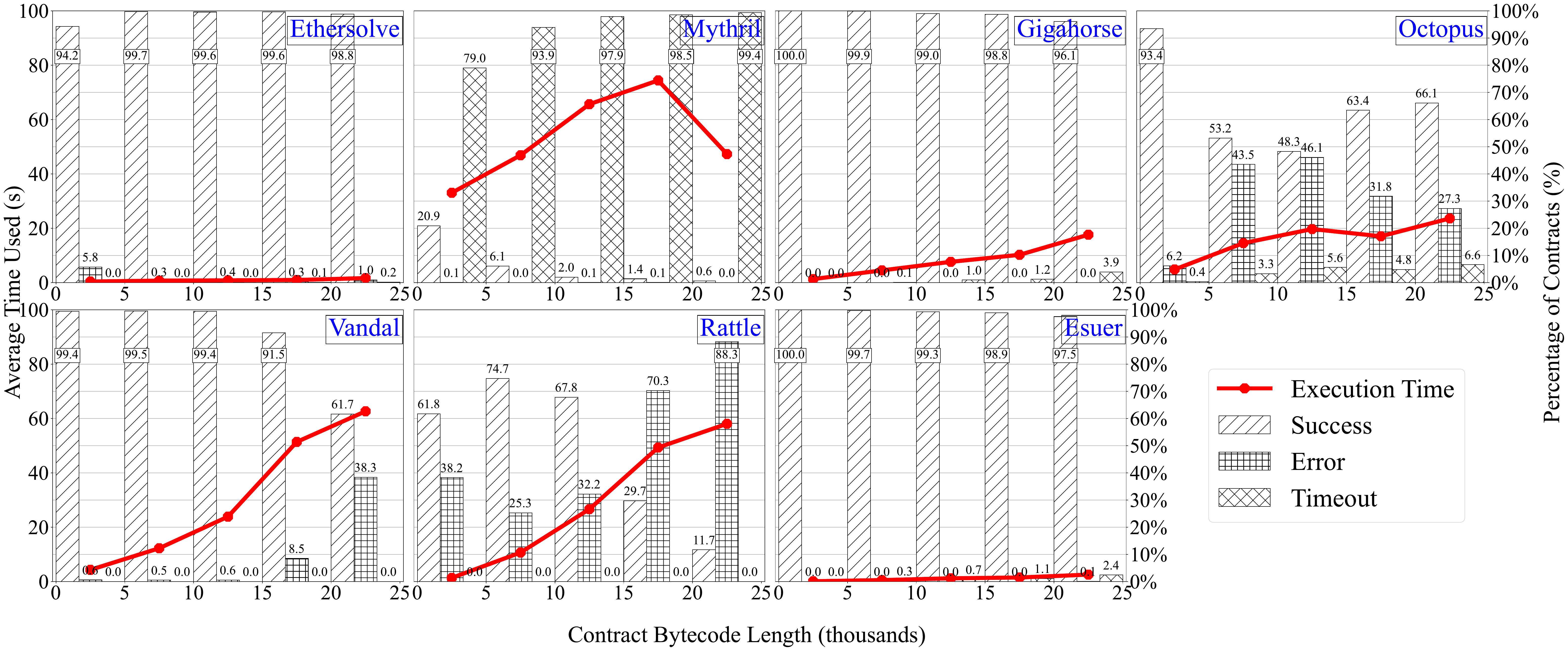}
    \vspace{-0.3cm}
    \caption{The execution results and execution time of smart contracts of different sizes. The red line represents the average execution time for contracts of different sizes (using the left y-axis), and the bars indicate the proportions of each execution state for contracts of different sizes (using the right y-axis).}
    \label{fig:scalability}
    \vspace{-0.3cm}
\end{figure*}

To answer RQ2, we evaluate performance of all tools on separate dockers with 16GB of memory and 4 cores, running 64-bit Ubuntu 20.04 LTS system. Each tool is given 120 seconds to analyze each smart contract and generate CFGs. If a tool does not output anything within the time limit, we consider it a timeout for that contract. Additionally, some tools may raise exceptions during processing for some contracts and fail to generate CFGs successfully, which we deem an error. Therefore, there are three possible results: success, error, and timeout. We also compare execution time of all tools and excludes their error and timeout cases for accuracy.
The performance of all tools is shown in Table~\ref{tab:performance}.

\noindent\textbf{Success rate}\tab
\system achieves a success rate of 99.25\%, which is the highest compared to other tools. The timeout cases of \system are due to dynamically changing reuse context, which causes \system to continuously generate clones for all 120 seconds. These cases can be resolved by combining dynamic analysis, which is outside the scope of this paper and will be considered in the future.
For Ethersolve and Octopus, most failures are caused by compiler conflicts because they rely on some magic code to analyze EVM bytecode. For Mythril, the symbolic execution and constraint solver make it time-consuming when handling most real-world smart contracts. As smart contracts become more complex, the necessity of static analysis for them is increasing. Gigahorse also has a high success rate, but for some cases, it only generates incomplete intermediate analysis results without any error information.
Vandal and Rattle use a backward strategy to construct CFGs from the leaf node to the root node, which makes them suffer from redundant data-flow dependencies caused by code reuse.

\noindent\textbf{Time}\tab
The average execution time of \system is 1.06 seconds, the second fastest among all tools. The fastest tool is Ethersolve, which is implemented in Java while the others are in Python.
However, as mentioned in \S\ref{subsec:precision}, the precision of Ethersolve is not that good because of its reuse-insensitive CFGs.
Although \system conducts code reuse identification, which Ethersolve does not, the average time of \system is only 0.2 seconds more than Ethersolve. This proves that the algorithm used by \system to identify code reuse is lightweight, and the overhead is negligible.
For Gigahorse, the average time is 7.40 seconds, nearly seven times of \system.
Although one may think it is just a few seconds, in practice, it is crucial for real-world malware and vulnerabilities detection.
The real-world attacks happen in real-time, so the time period between deploying malicious contracts or vulnerable contracts and attacks is limited for analysis.
This factor requires a high speed of tools to generate static primitives, especially when analyzing contracts in blockchains with a short block time such as BSC.
Therefore, the lightweight of \system makes it more practical in providing CFGs for large-scale static analysis towards real-world smart contracts.

\noindent\textbf{Scalability}\tab
We evaluate the scalability of all tools by grouping the success rate and execution time by the size of contracts.
We divide contract sizes into five levels, with an interval of 5,000 bytes.
The results are shown in Fig.~\ref{fig:scalability}.
For smart contracts of different sizes, \system all achieves high success rates and incurs small execution time.
The performance of \system is slightly influenced by the size of the analyzed contracts.
Compared with most other tools, for \system, contracts in larger sizes do not result in an exponential increase in execution time or a significant reduction in success rate.
The results demonstrate that \system has excellent scalability, enabling it to handle complex smart contracts of any size.

    \vspace{-0.1in}

\begin{framed}
    \vspace{-0.1in}
        \noindent
        \textbf{Answer to RQ2}\tab
        \system has the highest success rate of 99.25\% , the second fastest speed with an average execution time of 1.06s, and great scalability to handle real-world smart contracts.
        \vspace{-0.1in}
\end{framed}

\subsection{Effectiveness}

To answer RQ3, we implement two detectors based on \system and evaluate their ability to measure the effectiveness of \system. A case study is also provided to illustrate how \system assists in downstream static analysis.

\begin{table}[t]
    \centering
    \caption{Results of tx.origin and reentrancy detectors.}
    \label{tab:detectors}
    \vspace{-0.3cm}
    \footnotesize
    \resizebox{0.7\linewidth}{!}{%
    \begin{tabular}{ccccc} 
    \toprule
     & Tools & Precision & Recall & F1-score\\
    \midrule
    & \system & 99.93\% & 100\% & 99.97\% \\
    \rowcolor{mygray}
    \cellcolor{white} & Ethersolve & 99.93\% & 100\% & 99.97\% \\
    & Vandal & 98.86\% & 98.92\% & 98.89\% \\
    \rowcolor{mygray}
    \cellcolor{white} \multirow{-4}{*}{tx.origin} & Mythril & 75.27\% & 75.27\% & 75.27\% \\
    \midrule
    & \system & 99.41\% & 100\% & 99.67\% \\
    \rowcolor{mygray}
    \cellcolor{white} & Ethersolve & 84.26\% & 100\% & 91.04\% \\
     & Vandal & 79.39\% & 31.01\% & 43.23\% \\ 
    \rowcolor{mygray}
     \cellcolor{white} \multirow{-4}{*}{reentrancy} & Mythril & 42.46\% & 23.61\% & 29.07\% \\ 
    \bottomrule
    \end{tabular}
    }
\vspace{-0.3cm}
\end{table}

\subsubsection{Vulnerability detection}

We implement detectors built on top of \system to detect reentrancy vulnerabilities and tx.origin vulnerabilities.
We compare \system with Ethersolve, Vandal, and Mythril these three tools that have the capability to detect both reentrancy vulnerabilities and tx.origin vulnerabilities.
We use SolidiFI dataset, which contains smart contracts injected with vulnerabilities and ground truths about the locations of injected vulnerabilities.
All tools target EVM bytecode and report vulnerability locations in bytecode level, so we map ground truth from the source code to the bytecode, and evaluate the correctness of reports.
If the reported vulnerabilities are in the ground truth, they are TPs, otherwise, they are FPs.
If the vulnerabilities in the ground truth are not reported, they are FNs. 
We use the precision, recall, and F1-score metrics to measure the effectiveness of all tools.

The results are shown in Table~\ref{tab:detectors}.
\system outperforms all other tools.
For the tx.origin vulnerabilities whose pattern is simple, all tools perform better than reentrancy vulnerabilities.
\system and Ethersolve perform the best, with precision of 99.93\%, recall of 100\%, and F1-score of 99.97\%.
For the reentrancy vulnerabilities whose detection is complex, \system performs better than other tools.
That is because the detection of reentrancy vulnerabilities requires more precise control-flow information and data-flow information, which are provided by \system's reuse-sensitive analysis.

\begin{figure*}
    \centering
    \includegraphics[width = 0.9\linewidth]{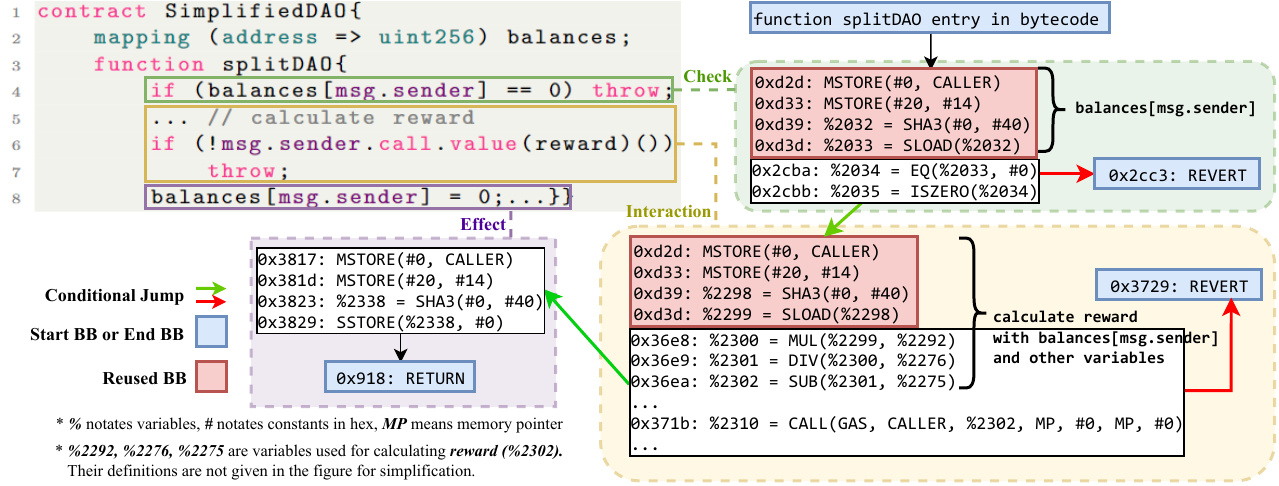}
    \vspace{-0.3cm}
    \caption{Simplified source code and generated CFG for the reentrancy vulnerability in the contract \textit{The DAO}.}
    \label{fig:caseStudy}
    \vspace{-0.3cm}
\end{figure*}


\subsubsection{Case Study}
\label{subsubsec:case_study}

We use the reentrancy vulnerability in the DAO attack~\cite{Dao} as the example, and simplify it for illustration.
The DAO attack is one of the most famous reentrancy attacks, causing a \$150 million loss in digital assets.
reentrancy occurs when the contract exposes external calls that can be hijacked by callers to call a function within the contract itself several times.
C-E-I (Check-Effect-Interaction)~\cite{cei1,cei2} is a security programming pattern in smart contracts to prevent reentrancy attacks.
It requires that external interactions should after checking preconditions and updating internal states.
The violation of C-E-I pattern can lead to reentrancy attacks by making use of the smart contract’s fallback mechanism.

For example, as the source code shown in Fig.~\ref{fig:caseStudy}, the function $splitDAO$ contains the operation of transferring Ethers.
In line 4, it first verifies whether the user’s balance ($balances[msg.sender]$) is more than 0 (Check phase).
Then the function calculates $reward$ that is the amount of transferred Ethers, with the user’s balance, which is simplified in line 5.
Line 6 conducts the transfer by $call.value$ (Interaction phase).
Following this, line 7 updates the user’s balance to 0 (Effect Phase).
This sequence violates the C-E-I pattern, which triggers reentrancy attacks.
This vulnerability arises in line 6, where the transfer awakes another contract's $fallback$ function.
In the $fallback$ function, attackers can repeatedly trigger the DAO's $splitDao$ function.
Because the balance has not been updated to 0 yet, the check in line 5 remains satisfied, facilitating further transfers.

The vulnerability in the bytecode is also shown in the Fig.~\ref{fig:caseStudy}.
We display key BBs in the \system-generated CFG and ignore others for illustration.
The BBs in the Check phase, Interaction phase, and Effect phase are framed with green, yellow, and purple rectangles, respectively.
The control flow in the CFG is the same as it in the source code, which also violates C-E-I pattern and can be detected by \system.
There is a BB whose offset is 0xd2d reused by both Check phase and Interaction phase, colored red in Fig.~\ref{fig:caseStudy}.
The reused BB is used to get the value of $balances[msg.sender]$, which is both used in check (line 4) and calculation of $reward$ (line 5), forming a fake loop which mixes Check phase and Interaction phase in the reuse-insensitive CFG .
With reuse-sensitive analysis, \system can identify the reused BB and transfer the fake loop into the origin control-flow structure, and identify the vulnerable C-I-E pattern clearly.

    \vspace{-0.1in}

\begin{framed}
    \vspace{-0.1in}
        \noindent
        \textbf{Answer to RQ3}\tab
        \system can improve the effectiveness of downstream static analysis, achieving a F1-score of 99.97\% to detect tx.origin vulnerabilities and a F1-score of 99.67\% to detect reentrancy vulnerabilities, outperforming other tools.
        \vspace{-0.1in}
\end{framed}

\section{Discussion}
\label{sec:discussion}

In this section, we discuss differences between code reuse and functions in EVM bytecode, and limitations of \system.

\noindent\textbf{Functions in EVM Bytecode}\tab
There are external functions and internal functions in smart contracts.
External functions exhibit consistent structures in EVM bytecode, whereas internal functions lack such regularity.
Therefore, most tool including our \system can extract external functions, which is unrelated to our main contributions and not mentioned before, and the focus of decompilers is to recover internal functions, which is out of our scope.
Someone may think code reuse is internal functions as they have similar behaviors.
However, actually, it is not.
Most reused code does not correspond to internal functions in source code.
As the example in \S\ref{subsubsec:case_study} shows, there do not exist internal functions in source code, while the BB 0xd2d is reused.
Code reuse is introduced by compilers, which makes EVM byteocde special.VIII

\noindent\textbf{Limitations}\tab
The limitations of \system include the following two aspects.
The first is that \system cannot handle jumps that use memory to store the jump operand or use calculated jump targets.
Processing these jumps is another challenge in constructing a solid and complete CFG and is out of the scope of this paper.
Fortunately, there are only a few cases today, and \system can handle most EVM bytecode.
A more comprehensive evaluation including memory space and calculable instructions can be considered to solve this problem, and can be a future direction to enhance our work.
The second is that \system cannot detect the reuse of BBs whose terminators end transactions such as STOP and REVERT.
That is because \system relies on jump targets in stacks to detect code reuse, but those end BBs do not have jump targets.
To address this problem, we make the following efforts.
For end BBs with multiple predecessors, \system regards it as a reused BB and clones it.
As shown in the evaluation in \S\ref{subsubsec:reuse_detection}, this method leads to some unnecessary clones of end BBs, but we think the influence to control flow and data flow is negligible and affordable.


\section{Related Work}
\label{sec:relatedwork}

\noindent\textbf{Program analyze tools for EVM bytecode}\tab
Various tools have been developed to analyze EVM bytecode.
To extract essential static primitives, Rattle~\cite{rattle}, Vandal~\cite{brent2018vandal}, and Ethersolve~\cite{contro2021ethersolve,pasqua2023enhancing} are developed.
They all construct reuse-insensitive CFGs for EVM bytecode, and are selected as comparison tools in the evaluation.
Our evaluation demonstrates that \system can provide more precise CFGs with reuse analysis.
To better understand EVM byteocde, several decompilers are developed using static analysis~\cite{grech2019gigahorse,grech2022elipmoc,zhou2018erays} or symbolic execution~\cite{eveem,etherscan_decompiler,suiche2017porosity}.
These works focus on recovering functions in the source code with different strategies.
Some of them~\cite{grech2019gigahorse,grech2022elipmoc,zhou2018erays} clone BBs that belong to different recovered functions, which mitigates the issue of code reuse.
However, our evaluation shows that they cannot eliminate this issue, leading to the flaws in their recovered source code.
In contrast to them, \system specifically targets code reuse and effectively eliminates this issue, which can assist in improving their results.
In addition, several works focus on detecting vulnerabilities in smart contracts.
These works use different techniques, such as static analysis~\cite{tsankov2018securify,brent2020ethainter} and symbolic execution~\cite{frank2020ethbmc,mythril,luu_making_2016,mossberg2019manticore}, with different scopes, such as certain types of vulnerabilities~\cite{bose2022sailfish,torres2018osiris,grech2018madmax,qian2023demystifying,wang2024efficiently} and cross-contract vulnerabilities~\cite{liao2022smartdagger,wang2024efficiently}.
They all rely on essential primitives during the analysis, which \system can provide.
Our evaluation shows that \system can enhance their effectiveness.

\noindent\textbf{Surveys on EVM bytecode tools}\tab
There are extensive studies about tools for EVM bytecode.
Chen et al.~\cite{chen2019large} evaluated the ability on control flow identification of six EVM bytecode tools.
The results reveal limitations in existing tools regarding control flow identification.
In this work, we point out the reason to the limitations, which is code reuse in EVM bytecode, and develop \system to address the issue.
Several works study the ability to detect vulnerabilities in smart contracts, with comprehensive benchmarks including generated buggy contracts~\cite{ghaleb2020effective}, annotated vulnerable smart contracts~\cite{durieux2020empirical} and DeFi attacks~\cite{chaliasos2023smart}, from various aspects including evaluating biases~\cite{ren2021empirical} and real-world impacts~\cite{chaliasos2024smart,li2024static,sendner2024large}.
Their results show that security tools require improvement to enhance their practical utility.
\system aims to provide precise essential primitives for these tools to improve their practical utility.

\noindent\textbf{Studies on code reuse}\tab
Extensive research has been conducted on code reuse in various domains, including smart contracts~\cite{sun2023demystifying,khan2022code,chen2021understanding}, blockchains~\cite{yi2022blockscope}, 
and so on~\cite{sun2014detecting,roy2007survey}.
These studies investigate code reuse resulting from developer practices.
In contrast, our work focuses on the code reuse attributable to the compiler.
While there have been studies on indirect jumps in other assembly languages~\cite{yakdan2015no, cifuentes1995decompilation}, leading to the reuse of assembly code, it is important to note that these cases originate from the source code and exhibit similar control flow semantics.
Instead, code reuse in EVM bytecode is introduced by the compiler itself, making it unpredictable and challenging to identify with conventional patterns.
To the best of our knowledge, we are the first to study the code reuse introduced by compiler instead of source code.

\section{Conclusion}
\label{sec:conclusion}

In this paper, we systematically study code reuse in EVM bytecode, which is the first to offer valuable insights into the challenges posed by compiler-introduced code reuse.
After that, we propose a lightweight algorithm for constructing reuse-sensitive CFGs, addressing these challenges and enhancing the precision, performance, and effectiveness of static analysis tools.
Our research provides a foundation for further exploration and advancement in the static analysis of EVM bytecode.
Future research can leverage these insights to develop more effective analysis tools to enhance the security of smart contracts.

\bibliographystyle{IEEEtran}
\bibliography{reference}


 




\vfill

\end{document}